\documentclass[apj]{emulateapj}
\usepackage{multirow}
\usepackage{longtable}
\usepackage{rotating}

\def\gtorder{\mathrel{\raise.3ex\hbox{$>$}\mkern-14mu
             \lower0.6ex\hbox{$\sim$}}}
\def\ltorder{\mathrel{\raise.3ex\hbox{$<$}\mkern-14mu
             \lower0.6ex\hbox{$\sim$}}}

\slugcomment{Accepted for publication in Icarus, Aug 17, 2012}

\shorttitle{Spectra and Spins of two Earth Grazing Asteroids}

\shortauthors{Polishook et al.}

\begin{document}

\title{Spectral and Spin Measurement of Two Small and Fast-Rotating Near-Earth Asteroids}
\author{D. Polishook\altaffilmark{1},
R.~P. Binzel\altaffilmark{1},
M. Lockhart \altaffilmark{1}
F.~E. DeMeo\altaffilmark{1},
W. Golisch\altaffilmark{2},
S.~J. Bus\altaffilmark{2},
A.~A.~S. Gulbis\altaffilmark{3},\altaffilmark{1},
}

\altaffiltext{1}{Department of Earth, Atmospheric, and Planetary Sciences, Massachusetts Institute of Technology, Cambridge, MA 02139, USA}
\altaffiltext{2}{Institute for Astronomy, 640 N. Aohoku Place, Hilo, HI 96720, United States}
\altaffiltext{3}{Southern African Large Telescope and the South African Astronomical Observatory, Cape Town, 7935}

\begin{abstract}
In May 2012 two asteroids made near-miss ``grazing" passes at distances of a few Earth-radii: 2012 KP24 passed at nine Earth-radii and 2012 KT42 at only three Earth-radii. The latter passed inside the orbital distance of geosynchronous satellites. From spectral and imaging measurements using NASA's 3-m Infrared Telescope Facility (IRTF), we deduce taxonomic, rotational, and physical properties. Their spectral characteristics are somewhat atypical among near-Earth asteroids: C-complex for 2012 KP24 and B-type for 2012 KT42, from which we interpret the albedos of both asteroids to be between 0.10 and 0.15 and effective diameters of $20\pm2$ and $6\pm1$ meters, respectively. Among B-type asteroids, the spectrum of 2012 KT42 is most similar to 3200 Phaethon and 4015 Wilson-Harrington. Not only are these among the smallest asteroids spectrally measured, we also find they are among the fastest-spinning: 2012 KP24 completes a rotation in $2.5008\pm0.0006$ minutes and 2012 KT42 rotates in $3.634\pm0.001$ minutes.


\end{abstract}


\section{Introduction}
\label{sec:Introduction}

Near-Earth asteroids (NEAs) that graze the Earth at distances of a few Earth-radii are a potential source of concern, but also present an opportunity to study small-sized asteroids in our immediate neighborhood. During a short window of opportunity, astronomers can measure the spectrum and brightness variation, from which the composition, size, rotation period, and shape can be interpreted.  

The results we report come from observations we conducted as part of a rapid response program on NASA's Infrared Telescope Facility (IRTF) on Mauna Kea, Hawaii.  Our program aims to measure the spectra of these Earth-grazing asteroids in near-infrared wavelengths. Due to the small sizes of these objects they are discovered only a few hours to a few days before encounter time forcing us to execute a target of opportunity program on short notice.

Here we report on two encounters, separated by a day only. The first object, 2012 KP24, crosses the orbits of the Earth and Mars and was discovered by the Catalina Sky Survey (Drake et al. 2009) five days before (geocentric) closest approach at 9 Earth-radii. The second object, 2012 KT42, was discovered by the Catalina Sky Survey only 23 hours before its closest approach at 3 Earth-radii, bringing it within the orbital distance of geosynchronous satellites (6.6 Earth-radii). The orbital elements of these two NEAs differ from one another and they do not have any physical relationship.

\section{Observations and Reduction}
\label{sec:observations}

Near-infrared spectroscopy was performed using SpeX, a 0.8-5.4 micron imager and spectrograph mounted on the IRTF (Rayner et al. 2003). Telescope tracking in the spectrograph slit was maintained using the MIT Optical Rapid Imaging System (MORIS; Gulbis et al. 2010) mounted on a side-facing exit window of SpeX. For these near-Earth asteroids, we simultaneously recorded 0.8 to 2.5 micron spectra and visible wavelength guide camera images. A set of exposures, 120 seconds each, were taken by SpeX while MORIS exposure time was two seconds. We used a slit width of 0.8 arcsec. No filter was used on MORIS in order to collect as many photons as possible. The weather on both nights was photometric with a typical seeing of 1.3 arcsec. For our 2012 May 28 (05:52 to 08:42) UT measurements of 2012 KP24, the tracking requirement was a ``modest" 1 arcsec/second rate relative to sidereal. For the following night's measurements of 2012 KT42 on 2012 May 29 (05:58 to 06:40) UT, non-sidereal tracking was required in excess of 10 arcsec/second. We prepared movies from the MORIS guiding images for both Earth-grazing encounters that are entertaining by their rapid motion relative to the background. The movies are presented in our website\footnote{http://smass.mit.edu/movie-caption.html}.

Processing of the raw SpeX images to obtain the final spectra followed the procedures outlined in Binzel et al. (2010) and DeMeo et al. (2009). The spectra were calibrated using measurements of the Solar analog stars 105-56, 107-684 and 107-998 taken before and after observing the asteroids that we identified from Landolt (1992) and have verified relative to solar analog stars 16 Cyg B and Hyades 64. Bias subtraction and twilight flat fielding were also part of the standard processing of the MORIS images. Flat fielding the MORIS images did not significantly change the results due to the short exposure time of the images, the high brightness of the asteroids, and the fact that the telescope was tracking at the asteroid rates, keeping them on the same pixels of the CCD. Apertures with 10-pixel radii (1.2 arcsec) were used for measuring the asteroid brightness with IRAF's phot function. Measurements contaminated by background stars were removed from the analysis. Because of the fast sky motion of the asteroids and the small field of view (1x1 square arcmin), other sources in the images appear as trails and could not be utilized as photometric reference stars for calibration, which made the lightcurves noisy. For studying the brightness variations, 2112 MORIS images were available for 2012 KP24, and 825 for 2012 KT42. The rapidly changing air mass, distance and phase angle relative to the Earth also created a linear brightness trend, most especially for 2012 KT42. This trend was removed prior to the period analysis, although we note that a long secondary period about a non-principal axis (Pravec et al. 2005) could also be present in that trend. We have no evidence for such a secondary period; we simply note that our observing window was too short to rule this out.

From the resulting photometric measurements, we found the synodic rotation period of the asteroids by fitting a second-order Fourier series to the data points following the methods described by Polishook et al. (2012). Using this technique, we scanned a range of frequencies and compared the $\chi^2$ value of each result to find the minimum value corresponding to the most likely period. The period uncertainty is set to $5\sigma$ above the minimal $\chi^2$.



\section{Results}
\label{sec:results}
\begin{figure}
\centerline{\includegraphics[width=8.5cm]{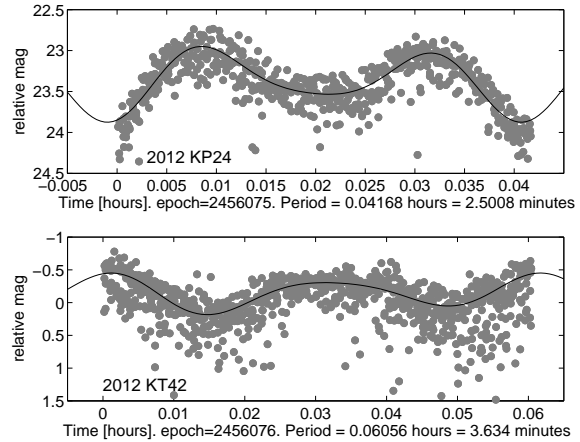}}
\caption{The lightcurve of 2012 KP24 folded by a period of 2.5008 minutes (upper panel) and the lightcurve of 2012 KT42 folded by a period of 3.634 minutes (lower panel).
\label{fig:FoldedLC}}
\end{figure}

We derived rotation periods of $2.5008\pm0.0006$ minutes for 2012 KP24 and $3.634\pm0.001$ minutes for 2012 KT42 (Figures~\ref{fig:FoldedLC}). Independent measurements performed by Warner et al. (2012) and L. Elenin\footnote{http://www.minorplanet.info/lightcurvedatabase.html} confirm our results. The lightcurve amplitudes are $0.9\pm0.2$ mag (2012 KP24) and $0.6\pm0.4$ mag (2012 KT42). Under the assumption of a triaxial shape ($a \geq b \geq c$), these lightcurve amplitudes correspond to minimal a/b axial ratios of $2.3\pm0.4$ (2012 KP24) and $1.7\pm0.6$ (2012 KT42).

\begin{figure}
\centerline{\includegraphics[width=8.5cm]{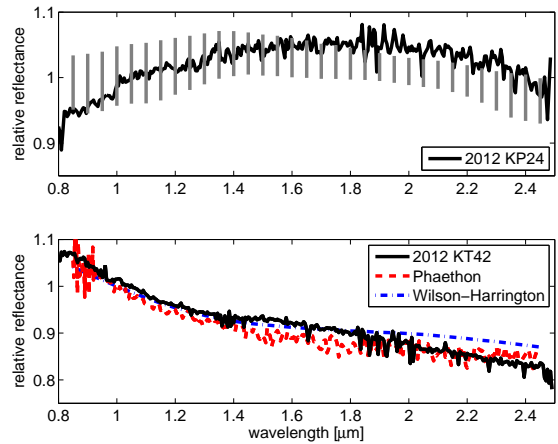}}
\caption{(top) Relative reflectance spectrum for 2012 KP24 compared with the range ($\pm1\sigma$ vertical bars) found within the C-complex. (bottom) Relative reflectance spectrum for 2012 KT42 compared with the B-type spectrum of asteroid 3200 Phaethon and a smoothed spectrum of 4015 Wilson-Harrington (Binzel et al. 2004).
\label{fig:Spectra}}
\end{figure}

The relative reflectance spectrum of 2012 KP24 (Figures~\ref{fig:Spectra}, top) is featureless and slightly convex, making it most closely consistent with asteroids in the C-complex defined by DeMeo et al. (2009). The absence of features and the unavailability of the visible spectral component inhibit any more specific class refinement beyond ``C-complex". For completeness we note its characteristics are only slightly beyond the bounds of the X-complex, further making a broad ``C-complex" conclusion the most conservative taxonomic assignment. On the other hand, the deeply negative spectral slope for 2012 KT42 (Figures~\ref{fig:Spectra}, bottom) is a unique characteristic of B-type asteroids. Among the eleven B-type asteroids in the SMASS database (Binzel et al. 2004), we note that 2012 KT42 is most similar to two of them: 3200 Phaethon and 4015 Wilson-Harrington. While spectral similarity does not necessarily imply any common relation, we note similar orbital inclinations (both near 2 degrees) for 2012 KT42 and Wilson-Harrington. Phaethon's much higher (22 degree) inclination more strongly precludes any obvious physical relationship.


\section{Discussion}
\label{sec:discussion}

The effective diameters of these objects are estimated from their absolute magnitudes {\it H} and inferred albedo values ${\it P_V}$. We use the absolute magnitude that appears in the Minor Planet Center (MPC): 26.4 mag for 2012 KP24 and 29 mag for 2012 KT42. These values, represent the magnitude of an asteroid in zero phase angle, are estimated by the MPC based on measurements taken on specific phase angles and a {\it G} slope of 0.15. To estimate the uncertainty of the absolute magnitude, {\it dH}, we calculate possible extreme values of the absolute magnitude by using extreme values of {\it G} slope 0 and 0.5 (Reddy et al. 2012), and by using the maximal phase angle the asteroid was observed at (70 degrees for 2012 KP24 and 6 degrees for 2012 KT42). This derives {\it dH} of +0.6 / -0.4 for 2012 KP24 and {\it dH} of +0.2 / -0.1 for 2012 KT42 (the error for 2012 KP24 is larger because it was observed at higher phase angle).

Based on the derived classes there is high likelihood the objects have relatively low albedos. However, the absence of any thermal flux (causing an upturn in the 2.0-2.5 micron region on the spectra of NEAs; Rivkin et al. 2005, Reddy et al. 2012) seems counter-indicative. This contradiction can be solved when assuming a regolith-free surface (typical of small-sized asteroids; Delbo et al. 2007) that increase the “beaming parameter”, $\eta$, of the thermal model, and reduce the thermal excess to insignificant values (see Fig. 4 in Rivkin et al. 2005). Therefore, we can infer only that the asteroid albedos are no lower than 0.10 and most likely for these types in the range 0.10 to 0.15. This range is consistent with the average albedo values of these taxonomies measured by the Wide-field Infrared Survey Explorer (WISE; Mainzer et al. 2011) and the value for 3200 Phaethon (Dumas et al. 1998). We estimate the maximum effective diameters, $D_{eff}$, using:
\begin{equation}
D_{eff} = \frac{1329}{\sqrt{P_V}}\times10^{-H/5}.
\label{eq:Deff}
\end{equation}
The uncertainty of the diameter is calculated using the extreme values of the absolute magnitude and the albedo. We derive maximum diameter values of $20\pm6$ meters for 2012 KP24 and $6\pm1$ meters for 2012 KT42. We deliberately avoid stating a minimal limit for the diameter, because there is a possibility that the albedos of the two asteroids are higher. However, such a minimal limit is still within a factor of two from the given value of the effective diameter.

\begin{figure*}
\centerline{\includegraphics[width=17cm]{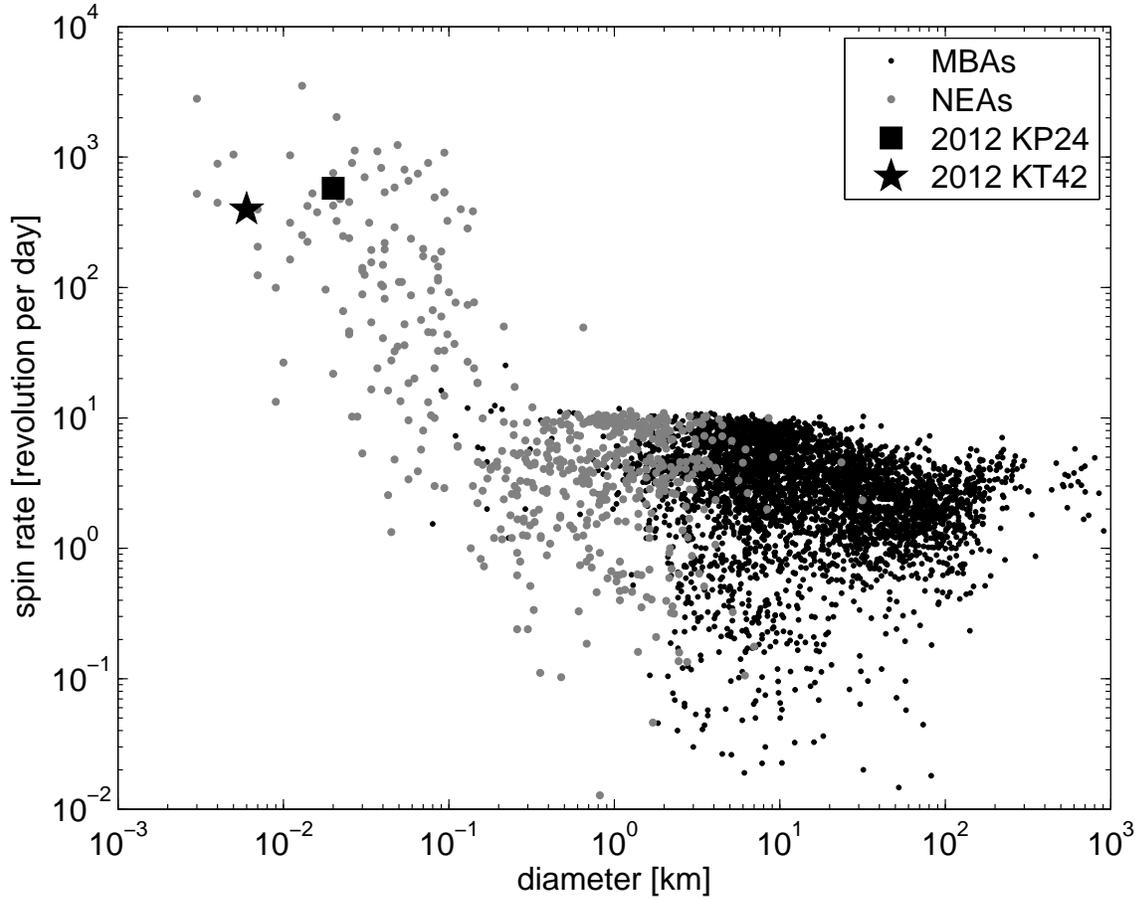}}
\caption{The diameter-spin rate diagram of asteroids marked with 2012 KP24 (square) and 2012 KT42 (star). The errors for these two asteroids are smaller than the symbols. The background population of asteroids is divided between Near-Earth asteroids (NEAs, grey dots) and main-belt asteroids (MBAs, black dots). Only secure rotation periods are plotted (quality code $U\geq2$). The data for the background population was taken from the Light Curve Data Base (LCDB), version of June 2012 (Warner et al. 2009).
\label{fig:DiamSpin}}
\end{figure*}

The sizes and lightcurve results make the two studied objects among the smallest and fastest-spinning measured asteroids. We mark the asteroids on the diagram of diameter vs. spin rate (Figures~\ref{fig:DiamSpin}) to display how the observation of such Earth grazing asteroids can help map the almost uncharted area of small, fast, and presumably intact asteroidal fragments that are typical of meteoroids capable of delivering samples. Excluding 2008 TC3, that crashed on Earth on October 2008 and its low signal-to-noise spectrum in the visible regime appears relatively flat (Jenniskens et al. 2009), none of the other seven objects among these asteroids ($D_{eff} \leq 200$ m, fast rotation and a known taxonomy) are of C-type or B-type as 2012 KP24 and 2012 KT42; implying that monolithic-structured asteroids can be formed from dark-type composition as well and they are not necessarily composed out of S-, X- or V-type materials.

Moreover, systematic measurement of these small and fast-rotating asteroids will determine if there is an observational bias against slow-rotators of the same size. A non-bias scenario will strengthen the idea that the Yarkovsky-O'Keefe-Radzievskii-Paddack (YORP) effect, which applies a thermal torque due to the re-emission of sunlight from an asymmetric surface (Rubincam 2000), has a fast and effective influence on such Near-Earth asteroids. Using MORIS for photometry while it is guiding the telescope to track an asteroid is an applicable method to answer this scientific question.

\acknowledgments

We thank Vishnu Reddy and Tomasz Kwiatkowski for their remarks on the manuscript. DP is grateful to the AXA research fund. RPB acknowledges NASA Near-Earth Object Observation program support through grant NNX10AG27G. Observations reported here were obtained at the Infrared Telescope Facility, which is operated by the University of Hawaii under Cooperative Agreement NCC 5-538 with the National Aeronautics and Space Administration, Science Mission Directorate, Planetary Astronomy Program. We thank J. Rayner and A. Tokunaga for their help with the IRTF. We appreciate the Hawaiian people for allowing us to use their sacred land for astronomy.

\end{document}